\documentclass[twocolumn]{aastex631}

\usepackage{amsmath}
\usepackage{amssymb}
\usepackage{natbib}
\usepackage{CJKutf8}
\usepackage{enumitem}

\usepackage{savesym}
\savesymbol{tablenum}
\usepackage{siunitx}
\newcommand{\Msun}{M_{\odot}}

\DeclareSIUnit\dyne{dyn}
\DeclareSIUnit\year{yr}

\newcommand{\Mpri}{M}
\newcommand{\Rpri}{R}

\newcommand{\mratio}{q}

\newcommand{\vpos}{\mathbf{r}}

\newcommand{\ver}{\mathbf{e}_{r}}
\newcommand{\vet}{\mathbf{e}_{\vartheta}}
\newcommand{\vep}{\mathbf{e}_{\varphi}}

\newcommand{\Oorb}{\Omega_{\rm orb}}

\newcommand{\epstide}{\varepsilon_{\rm T}}

\newcommand{\sigmk}{\sigma_{m,k}}

\newcommand{\Orot}{\Omega_{\rm rot}}

\newcommand{\frot}{f_{\rm rot}}
\newcommand{\forb}{f_{\rm orb}}
\newcommand{\fps}{f_{\rm ps}}

\newcommand{\rs}{r_{\rm s}}

\newcommand{\potpri}{\Phi}

\newcommand{\pottide}[1][0]{\Phi_{{\rm T}\if#10\else;#1\fi}}
\newcommand{\pottot}[1][0]{\Psi\if#10\else_{#1}\fi}

\newcommand{\vxi}{\boldsymbol{\xi}}

\newcommand{\angtime}[1][0]{H_{\if#10\else#1\fi}}

\newcommand{\txir}[1][0]{\tilde{\xi}_{{\rm r}\if#10\else;#1\fi}}
\newcommand{\txih}[1][0]{\tilde{\xi}_{{\rm h}\if#10\else;#1\fi}}

\newcommand{\tf}[1][0]{\tilde{f}_{\if#10\else#1\fi}}
\newcommand{\tP}[1][0]{\tilde{P}_{\if#10\else#1\fi}}

\newcommand{\tpotpri}[1][0]{\tilde{\potpri}_{\if#10\else#1\fi}}
\newcommand{\tpottot}[1][0]{\tilde{\Psi}_{\if#10\else#1\fi}}
\newcommand{\tpottide}[1][0]{\tilde{\Phi}_{{\rm T}\if#10\else;#1\fi}}

\newcommand{\trho}[1][0]{\tilde{\rho}_{\if#10\else#1\fi}}
\newcommand{\tS}[1][0]{\tilde{S}_{\if#10\else#1\fi}}
\newcommand{\tT}[1][0]{\tilde{T}_{\if#10\else#1\fi}}

\newcommand{\tFradr}[1][0]{\tilde{F}_{{\rm rad},r\if#10\else;#1\fi}}

\newcommand{\teps}[1][0]{\tilde{\epsilon}_{\if#10\else#1\fi}}
\newcommand{\tkap}[1][0]{\tilde{\kappa}_{\if#10\else#1\fi}}
\newcommand{\tLrad}[1][0]{\tilde{L}_{{\rm R}\if#10\else;#1\fi}}

\newcommand{\ecoeff}[1][0]{A_{\if#10\else#1\fi}}
\newcommand{\ezcoeff}[1][0]{A^{0}_{\if#10\else#1\fi}}

\newcommand{\profile}[1][0]{\Delta_{\if#10\else#1\fi}}

\newcommand{\cbar}[1][0]{\bar{c}_{\if#10\else#1\fi}}
\newcommand{\Fbar}[1][0]{\bar{F}_{\if#10\else#1\fi}}
\newcommand{\Gbar}[2][0]{\bar{G}^{(#2)}_{\if#10\else#1\fi}}

\newcommand{\kap}[1][0]{\kappa_{\if#10\else#1\fi}}

\newcommand{\torque}{\mathcal{T}}

\newcommand{\torsec}{\torque_{\rm sec}}

\newcommand{\Wbar}[1][0]{\bar{W}_{\if#10\else#1\fi}}
\newcommand{\Deltabar}[1][0]{\bar{\Delta}_{\if#10\else#1\fi}}

\newcommand{\engy}[1][0]{\mathcal{E}_{\if#10\else#1\fi}}
\newcommand{\work}[1][0]{\mathcal{W}_{\if#10\else#1\fi}}

\newcommand{\gamlmk}{\gamma_{\ell,m,k}}

\newcommand{\omegaf}{\omega}

\newcommand{\gmode}[1]{{\rm g}_{#1}}

\newcommand{\hxir}[1][0]{\hat{\xi}_{{\rm r}\if#10\else;#1\fi}}
\newcommand{\hxih}[1][0]{\hat{\xi}_{{\rm h}\if#10\else;#1\fi}}

\newcommand{\hpotpri}[1][0]{\hat{\potpri}_{\if#10\else#1\fi}}

\newcommand{\hsignl}{\hat{\sigma}_{n,\ell}}
\newcommand{\hgamnl}{\hat{\gamma}_{n,\ell}}

\newcommand{\imag}{\operatorname{Im}}

\newcommand{\ii}{\mathrm{i}}

\newcommand{\diff}[1]{\operatorname{d}\!{#1}}
\newcommand{\deriv}[3][0]{\frac{\operatorname{d}\if#10\else^{#1}\!\fi\!{#2}}{\operatorname{d}\!{#3}\if#10\else^{#1}\fi}}
\newcommand{\sderiv}[3][0]{\operatorname{d}\if#10\else^{#1}\fi\!{#2}/\operatorname{d}\!{#3}\if#10\else^{#1}\fi}
\newcommand{\pderiv}[3][0]{\frac{\partial\if#10\else^{#1}\!\fi{#2}}{\partial{#3}\if#10\else^{#1}\fi}}
\newcommand{\spderiv}[3][0]{\partial\if#10\else^{#1}\fi{#2}/\partial{#3}\if#10\else^{#1}\fi}

\newcommand{\gyre}{GYRE}
\newcommand{\mesa}{MESA}

\newcommand{\xgyretides}{\texttt{gyre\_tides}}

\begin{document}

\title{Discrepant Approaches to Modeling Stellar Tides, and the
  Blurring of Pseudosynchronization}

\author[0000-0002-2522-8605]{R. H. D. Townsend}
\affiliation{Department of Astronomy, University of Wisconsin-Madison, 475 N Charter St, Madison, WI 53706, USA}
\author[0000-0001-9037-6180]{M. Sun}
\affiliation{Center for Interdisciplinary Exploration and Research in Astrophysics (CIERA), Northwestern University, 1800 Sherman Ave, Evanston, IL 60201, USA}

\begin{abstract}
We examine the reasons for discrepancies between two alternative
approaches to modeling small-amplitude tides in binary systems. The
`direct solution' (DS) approach solves the governing differential
equations and boundary conditions directly, while the `modal
decomposition' (MD) approach relies on a normal-mode
expansion. Applied to a model for the primary star in the heartbeat
system KOI-54, the two approaches predict quite different behavior of
the secular tidal torque. The MD approach exhibits the
pseudosynchronization phenomenon, where the torque due to the
equilibrium tide changes sign at a single, well-defined and
theoretically predicted stellar rotation rate. The DS approach instead
shows `blurred' pseudosynchronization, where positive and negative
torques intermingle over a range of rotation rates.

We trace a major source of these differences to an incorrect damping
coefficient in the profile functions describing the frequency
dependence of the MD expansion coefficients. With this error corrected
some differences between the approaches remain; however, both are in
agreement that pseudosynchronization is blurred in the KOI-54
system. Our findings generalize to any type of star for which the
tidal damping depends explicitly or implicitly on the forcing
frequency.
\end{abstract}

\keywords{Binary stars (154) --- Tides (1702) --- Stellar oscillations (1617) --- Orbital evolution (1178) --- Theoretical techniques (2093)}

\section{Introduction} \label{s:intro}

\citet[][hereafter S23]{Sun:2023} introduce new functionality in the
\gyre\ oscillation code
\citep{Townsend:2013,Townsend:2018,Goldstein:2020} for modeling
small-amplitude tides in binary systems. This functionality is
implemented using a `direct solution' (DS) approach, in which the
differential equations and boundary conditions governing the radial
dependence of tidal perturbations are solved directly as a two-point
boundary value problem. To validate their implementation, S23 compare
it against an alternative `modal decomposition' (MD) approach popular
in the literature \citep[see,
  e.g.,][]{Press:1977,Kumar:1995,Lai:1997,Fuller:2012,Burkart:2012},
where perturbations are expanded as a superposition of the star's
free-oscillation normal modes. An unexpected finding is that the
secular tidal torques calculated using DS and MD differ significantly,
casting doubt on whether the two approaches are truly equivalent.

In this paper we delve into the reasons for these differences. We
focus on a model for the $\SI{2.32}{\Msun}$ primary star of the highly
eccentric ($e \approx 0.8$) heartbeat system KOI-54, constructed by
S23 using the \mesa\ stellar evolution code
\citep{Paxton:2011,Paxton:2013,Paxton:2015,Paxton:2018,Paxton:2019,Jermyn:2022}. This
model, about a third of the way through its main-sequence lifetime,
has a convective core and a radiative envelope; as such, the principal
tidal damping mechanism in the star is radiative dissipation. Most of
our analysis is aimed at stars for which this is likewise the case,
although we briefly discuss the relevance of our findings for other
types of star.

The following section begins with a demonstration of the
problem. In Section~\ref{s:diag} we track down the reason for the
differences between the MD and DS approaches, and in Section~\ref{s:fix} we propose a
fix to MD to resolve the issue. We reflect on the 
blurring of pseudosynchronization in
Section~\ref{s:pseudo}, and then conclude in Section~\ref{s:discuss}
with a summary and discussion of our findings.


\section{Demonstrating the Problem} \label{e:demo}

\begin{figure*}
  \includegraphics{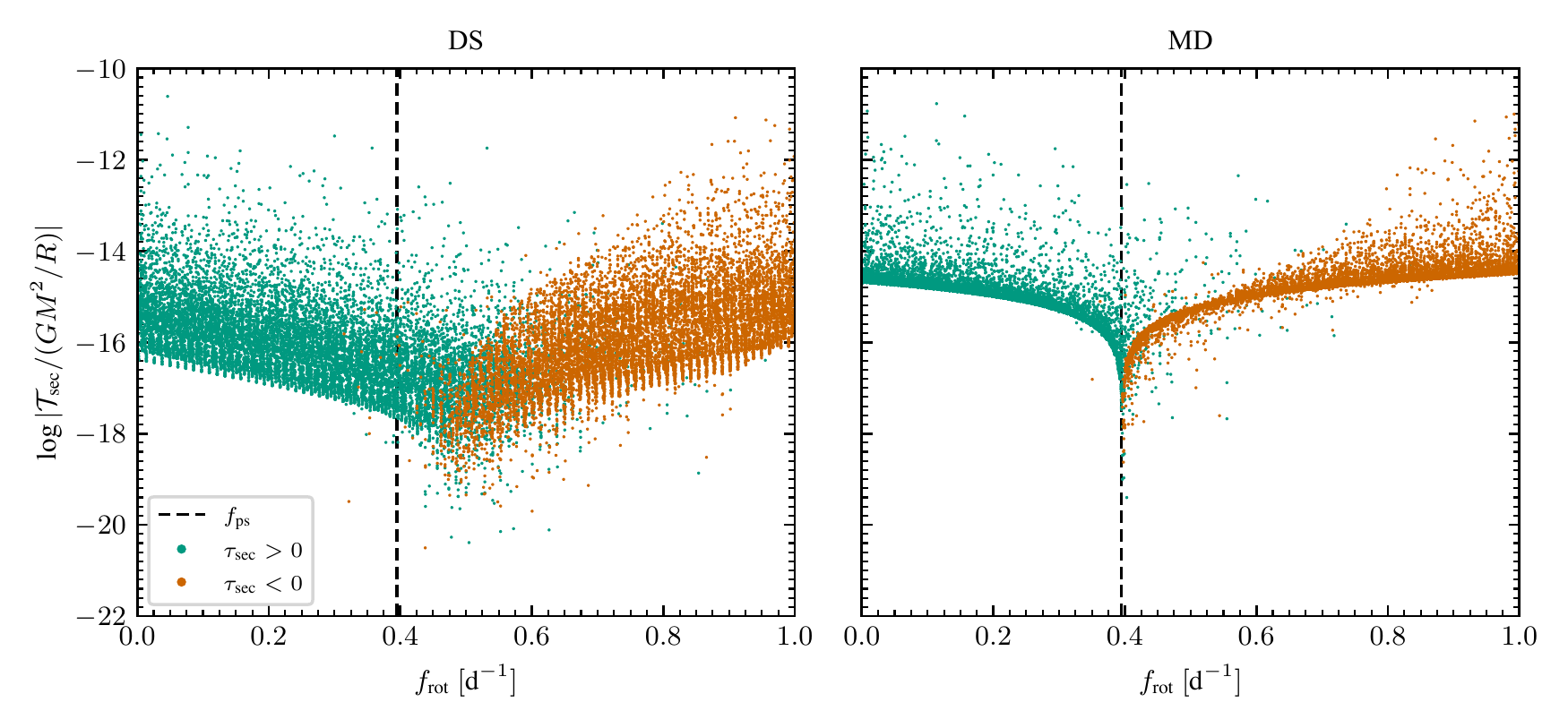}
  \caption{Secular torque $\torsec$ plotted against stellar rotation
    frequency $\frot$ for the KOI-54 primary model, as calculated
    using the DS approach (left) and the MD approach (right). In both
    panels, the vertical dashed line indicates the pseudosynchronous
    rotation rate $\fps$ given by equation~(\ref{e:pseudo}).}
  \label{f:torque}
\end{figure*}

Figure~\ref{f:torque} demonstrates the differing predictions of DS and
MD. Reprising Fig.~8 of S23, it plots the secular torque $\torsec$ due
to the quadrupole tide as a function of stellar rotation frequency
$\frot$, for the KOI-54 primary model. The two panels illustrate the
outcome of using the DS approach (left; via \xgyretides) and the MD
approach (right; via the formalism described in the following
section). The torque is an extremely sensitive function of rotation
frequency, varying by orders of magnitude as different
free-oscillation modes are Doppler-shifted in and out of resonance;
therefore, rather than attempting to plot continuous curves we sample
the torque at 25,000 rotation frequencies randomly distributed in the
interval $0 \leq \frot/\si{\per\day} \leq 1$, and plot these data as
discrete points in the figure.

The two panels agree in their general characteristics: torques are
mostly positive at slow rotation rates, mostly negative at rapid
rotation rates, and exhibit a lower envelope corresponding to the
equilibrium tide (with departures from this envelope occurring near
modal resonances.) However, the transition between the slow- and
rapid-rotation limits occurs differently in each case. With MD, the
equilibrium tidal torque passes through zero at $\frot =
\SI{0.395}{\per\day}$, coinciding with the pseudosynchronous rate
\begin{equation} \label{e:pseudo}
  \fps \equiv \forb \frac{1 + (15/2)e^{2} + (45/8)e^{4} +
    (5/16)e^{6}}{[1 + 3e^{2} + (3/8)e^4](1 - e)^{3/2}}
\end{equation}
predicted by \citet{Hut:1981}; here, $\forb$ is the orbital
frequency. Within DS, however, the switch from positive to negative
torque occurs over a range $1.1 \la \frot/\fps \la 1.4$ of rotation
frequencies, with equilibrium torques of opposite sign intermingling
across this interval. We dub this `blurred'
pseudosynchronization. Away from the interval, the torque is
one-to-two orders of magnitude smaller than in the MD case.


\section{Diagnosing the Problem} \label{s:diag}

\subsection{Preliminaries}

To explore the reasons for the discrepancies shown in
Fig.~\ref{f:torque}, we first recapitulate some of the formalism from
S23. The displacement perturbation vector $\vxi$ of a star, responding
to strictly periodic tidal forcing by a companion, can be expressed as
\begin{multline} \label{e:sol-xi}
  \vxi(\vpos;t) =
  \sum_{\ell,m,k}
  \left[
    \txir[\ell,m,k](r) \, \ver \phantom{\left( \vet \, \pderiv{}{\vartheta} \right)}
  \right. \\
  \left. \mbox{} + 
    \txih[\ell,m,k](r) \left( \vet \, \pderiv{}{\vartheta} + \frac{\vep}{\sin\vartheta} \, \pderiv{}{\varphi}\right)
    \right] \\
  \mbox{} \times Y^{m}_{\ell}(\vartheta,\varphi) \, \exp(-\ii k \Oorb t).
\end{multline}
Here, $\ver$, $\vet$ and $\vep$ are the unit basis vectors in the
radial ($r$), polar ($\vartheta$), and azimuthal ($\varphi$)
directions, respectively; $Y^{m}_{\ell}$ is the spherical harmonic
with harmonic degree $\ell$ and azimuthal order $m$; $\Oorb \equiv
2\pi \forb$ is the orbital angular frequency; and the summations
extend over $\ell \in [2,\infty]$, $m \in [-\ell,\ell]$ and
Fourier-series index $k \in [-\infty,\infty]$. The Eulerian
perturbation to the star's self-gravitational potential is likewise
given by
\begin{equation} \label{e:sol-f}
  \potpri'(\vpos;t) =
  \sum_{\ell,m,k}
  \tpotpri[\ell,m,k]'(r) \, Y^{m}_{\ell}(\vartheta,\varphi) \, \exp(-\ii k \Oorb t),
\end{equation}
and similar expressions exist for the perturbations to other scalar
quantities.

The functions $\txir[\ell,m,k]$, $\txih[\ell,m,k]$ and
$\tpotpri[\ell,m,k]'$ appearing in these expressions encapsulate the
radial dependence of the response to the partial tidal potential
$\pottide[\ell,m,k](\vpos;t)$ (see equation~9 of S23). In the DS
approach these functions are found by solving the two-point boundary
value problem comprising the tidal equations and boundary conditions,
as detailed in Appendix E of S23. In the MD approach, the functions
are instead expanded as a superposition of the star's adiabatic
normal-mode eigenfunctions. Thus, for instance, $\tpotpri[\ell,m,k]'$ is
expressed as
\begin{equation} \label{e:md-exp}
  \tpotpri[\ell,m,k]'(r) = \sum_{n} \ecoeff[n,\ell,m,k] \, \hpotpri[n,\ell]'(r)
\end{equation}
where $\hpotpri[n,\ell]'$ is the self-gravitational potential
perturbation eigenfunction associated with the normal mode of harmonic
degree $\ell$ and suitably-defined\footnote{Although no general
algorithm exists for assigning each $\ell \geq 2$ mode a unique radial
order, this isn't a problem as long as \emph{all} modes with the same
$\ell$ are included in the summation.} radial order
$n$. \citet{Burkart:2012} provide an expression for the expansion
coefficients $\ecoeff[n,\ell,m,k]$ (see their equation 7); in the
present notation this can be written
\begin{equation} \label{e:coeff}
  \ecoeff[n,\ell,m,k] = \frac{2 \epstide \, Q_{n,\ell} \, \cbar[\ell,m,k] \, \profile[n,\ell,m,k]}{E_{n,\ell}}.
\end{equation}
Here, the overlap integral $Q_{n,\ell}$ and normalized energy
$E_{n,\ell}$ are defined in equations~(9) and~(10), respectively, of
\citet{Burkart:2012}. The profile functions
\begin{equation} \label{e:profile}
  \profile[n,\ell,m,k] \equiv \frac{\sigmk^{2}}{(\hsignl^{2} - \sigmk^{2}) - 2 \ii \hgamnl \sigmk}
\end{equation}
encapsulate the frequency dependence of the expansion coefficients,
with $\hsignl$ the mode eigenfrequencies, $\hgamnl$ the damping
coefficients\footnote{\citet{Burkart:2012} obtain these damping
coefficients using a quasi-adiabatic integral (see their equation
14). In the present work, however, we follow S23 and determine
$\hgamnl$ from the imaginary part of \emph{non-adiabatic}
eigenfrequencies.}, and
\begin{equation} \label{e:sigmk}
  \sigmk \equiv k \Oorb - m \Orot
\end{equation}
the angular frequency of tidal forcing in the frame co-rotating with
the primary\footnote{The second term on the right-hand side of
equation~(\ref{e:sigmk}) is the Doppler shift arising when
transforming from an inertial frame to the co-rotating frame. In the
present analysis, as in S23, we neglect the effects of the centrifugal
and Coriolis forces due to rotation; therefore, this term is the sole
place in our formalism where $\Orot$ appears.}. Other symbols, here
and subsequently, follow the definitions given in S23.

With the radial functions determined, the secular tidal torque plotted
in Fig.~\ref{f:torque} follows as
\begin{multline}
  \torsec = 
  4 \Oorb \sqrt{\frac{G \Mpri^{3} \mratio^{4}}{1 + \mratio}} \, a^{1/2}
  \sum_{\ell,m,k \geq 0}
  \left( \frac{\Rpri}{a} \right)^{\ell+3} \,
  \left( \frac{\rs}{\Rpri} \right)^{\ell+1} \\
  \times \kap[\ell,m,k] \,
  \imag(\Fbar[\ell,m,k]) \,
  \Gbar[\ell,m,k]{4}
\end{multline}
(from equation 25 of S23; the summation is now over non-negative $k$). The normalized response functions
\begin{equation}
  \Fbar[\ell,m,k] \equiv - \frac{1}{2} \frac{\Rpri}{G\Mpri} \frac{\tpotpri[\ell,m,k]'(\rs)}{\cbar[\ell,m,k] \epstide},
\end{equation}
with $\rs$ the surface radius of the primary, are proportional to the tidal Love numbers $k_{\ell}$. These functions also appear in corresponding expressions for the secular
rates-of-change of the orbital elements (see, e.g., equation 23 of
S23), and the rate-of-work done by the tide (see Section~\ref{s:fix}); 
they are therefore an important output of any numerical tidal
calculation.

\subsection{Response Functions}

\begin{figure}
  \includegraphics{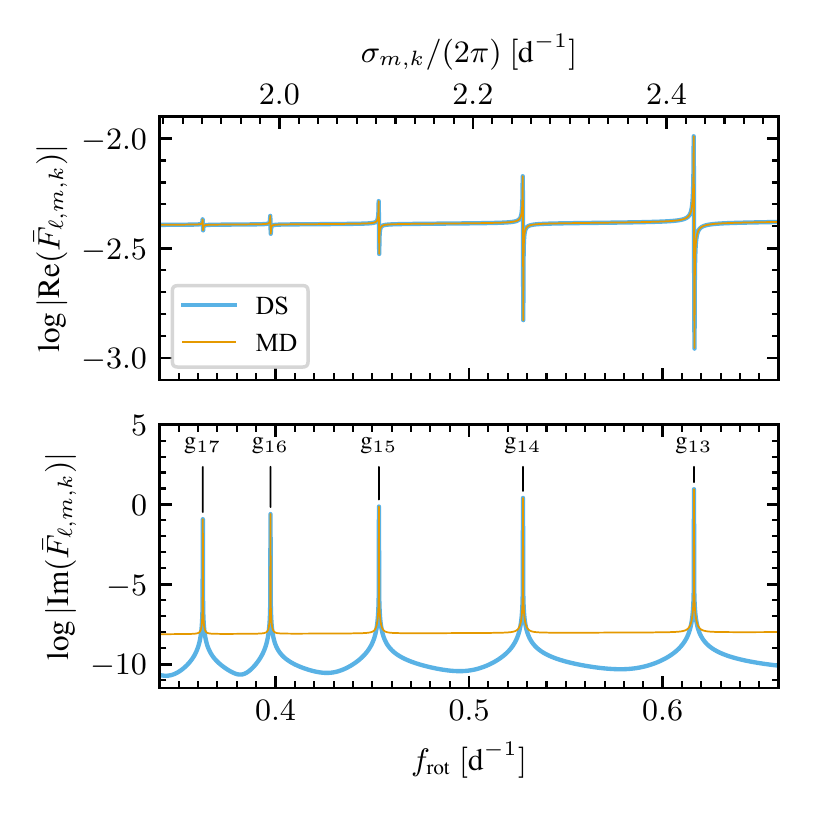}
  \caption{Normalized response function $\Fbar[\ell,m,k]$ plotted
    against stellar rotation frequency $\frot$, for the $\{\ell,m,k\}
    = \{2,-2,50\}$ partial tide of the KOI-54 primary model. The real
    (imaginary) part of the function is shown in the upper (lower)
    panel, and separate curves are plotted for the DS and MD
    approaches. Peaks are labeled by the resonant mode’s
    classification within the Eckart-Osacki-Scuflaire scheme. The
    upper axis scale shows the tidal forcing angular frequency
    $\sigmk$ (equation~\ref{e:sigmk}).}
  \label{f:resp-func}
\end{figure}

We begin our diagnosis by comparing the normalized response functions predicted
by the DS and MD approaches, for forcing by the $\{\ell,m,k\} = \{2,
-2, 50\}$ partial tidal potential in the KOI-54 primary
model. Figure~\ref{f:resp-func} plots the real and imaginary parts of
$\Fbar[\ell,m,k]$ as a function of rotation frequency, following each
approach. In both panels the distinct peaks corresponding to
resonances with the star’s $\ell=2$ free-oscillation modes, and are
labeled with the resonant mode's classification within the
Eckart-Osaki-Scuflaire scheme \citep[see, e.g.,][]{Unno:1989}.

In the vicinity of the peaks both approaches appear to agree
well. However, between the peaks the $\imag(\Fbar[\ell,m,k])$ data
deviate significantly; the MD values are up to two orders of magnitude
larger than the DS ones, and show a much weaker frequency
dependence. These deviations are responsible for the
discrepancies seen in Fig.~\ref{f:torque}.

\subsection{Profile Functions}

The deviations in Fig.~\ref{f:resp-func} are frequency-dependent,
and thus can only arise via the profile
functions~(\ref{e:profile}). To shed light on what might be amiss with
these functions, we apply the MD formalism to the analogous but
simpler problem of forced, damped transverse waves on a stretched
string. Where appropriate, we draw direct parallels to the MD
derivation by \citet{Kumar:1995}.

Let $\tau$ be the tension and $\mu$ the mass per unit length
of the string; the equation of motion for small transverse
displacements $u(x,t)$ is then
\begin{equation} \label{e:string-wave}
  \pderiv[2]{u}{t} = c^{2} \pderiv[2]{u}{x} - 2 \gamma \pderiv{u}{t},
\end{equation}
where $c \equiv \sqrt{\tau/\mu}$ is the wave speed, and the second
term on the right-hand side models the damping as a velocity-dependent
(Stokes) drag force governed by the coefficient $\gamma$. We assume
the string is clamped at one end and subject to a transverse force
$f(t)$ at the other, leading to the boundary conditions
\begin{equation}
  u = 0
\end{equation}
at the clamped end ($x=0$) and
\begin{equation} \label{e:bc-inhom}
  \tau \pderiv{u}{x} = f
\end{equation}
at the forced end ($x=L$, with $L$ the string length).

To solve the stretched-string problem, we first reduce the
non-homogeneous boundary condition~(\ref{e:bc-inhom}) to a homogeneous
one using the substitution
\begin{equation} \label{e:hom-trans}
u(x,t) = v(x,t) + \frac{f x}{\tau}.
\end{equation}
The wave equation~(\ref{e:string-wave}) then becomes
\begin{equation} \label{e:string-wave-inhom}
  \pderiv[2]{v}{t} = c^{2} \pderiv[2]{v}{x} - 2 \gamma \pderiv{v}{t} -
  \frac{x}{\tau} \left( \deriv[2]{f}{t} + 2 \gamma \deriv{f}{t} \right),
\end{equation}
and the boundary conditions are
\begin{equation}
  v = 0
\end{equation}
at $x=0$ and
\begin{equation}
\pderiv{v}{x} = 0
\end{equation}
at $x=L$.

In the spirit of the MD approach, we assume solutions $v(x,t)$ can be expressed
as a superposition of the spatial eigenfunctions that result from
solving the homogeneous, undamped counterpart to the wave equation~(\ref{e:string-wave-inhom}). Thus,
\begin{equation} \label{e:eig-exp}
  v(x,t) = \sum_{n=1}^{\infty} T_{n}(t) \sin(\omega_{n} x / c)
\end{equation}
\citep[cf. equation 1 of][]{Kumar:1995}, where
\begin{equation}
  \omega_{n} = \left( n - \tfrac{1}{2} \right) \frac{\pi c}{L} \qquad (n = 1,2,\ldots)
\end{equation}
are the eigenvalues. Substituting the trial solutions~(\ref{e:eig-exp}) into
equation~(\ref{e:string-wave-inhom}), and leveraging the orthogonality
of the eigenfunctions, leads to an ordinary differential equation for
each of the temporal coefficients $T_{n}(t)$:
\begin{equation} \label{e:harm-osc}
  \deriv[2]{T_{n}}{t} + 2 \gamma \deriv{T_{n}}{t} + \omega_{n}^{2} T_{n} = F_{n}(t),
\end{equation}
\citep[cf. equation~2 of][]{Kumar:1995}, where
\begin{equation}
  F_{n}(t) =
  (-1)^{n} \frac{2c^{2}}{\omega_{n}^{2} \tau L} \left( \deriv[2]{f}{t} + 2 \gamma \deriv{f}{t} \right).
\end{equation}
Equation~(\ref{e:harm-osc}) has the familiar form of the second-order
ordinary differential equation governing the motion of a forced,
damped harmonic oscillator. For periodic external forcing
\begin{equation}
  f(t) = f_{0} \exp(-\ii \omegaf t),
\end{equation}
steady-state solutions follow as
\begin{multline}
  T_{n}(t) = (-1)^{n+1} \frac{2 f_{0} c^2}{\omega_{n}^{2} \tau L} \frac{\omegaf^{2} + 2 \ii \gamma \omegaf}{(\omega_{n}^{2} - \omegaf^{2}) - 2 \ii \gamma \omegaf} \\
  \mbox{} \times \exp(-\ii \omegaf t).
\end{multline}
Combining this with equations~(\ref{e:hom-trans})
and~(\ref{e:eig-exp}), the steady-state solution to the original
problem is
\begin{multline}
  u(x,t) = \frac{2 f_{0} c^{2}}{\omegaf^{ 2} \tau L} \sum_{n=1}^{\infty} (-1)^{n+1} \profile[n] \, \sin(\omega_{n} x/c) \\
  \mbox{} \times \exp(-\ii \omegaf t),
\end{multline}
where the profile functions are
\begin{equation}
  \profile[n] \equiv \frac{\omegaf^{2}}{(\omega_{n}^{2} - \omegaf^{2}) - 2 \ii \gamma \omegaf}.
\end{equation}
\citep[cf. equation 10 of][]{Kumar:1995}. Comparing this result
against the profile functions~(\ref{e:profile}) arising in the MD
approach for stellar tides, a noteworthy difference is that the
damping coefficient in the latter is subscripted by the radial order
$n$. We maintain that this is incorrect --- as our analysis here of
the stretched-string problem demonstrates, $\gamma$ should be independent of the summation index arising in the modal decomposition.


\section{Fixing the Problem} \label{s:fix}

\begin{figure}
  \includegraphics{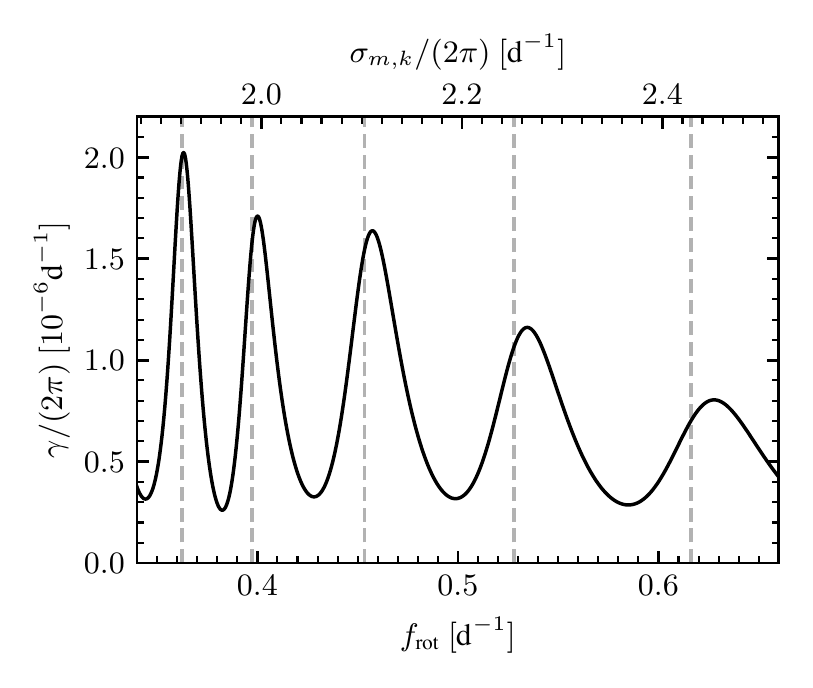}
  \caption{Damping coefficient $\gamlmk$ plotted against stellar
    rotation frequency $\frot$, for the
    $\{\ell,m,k\}=\{2,-2,50\}$ partial tide of the KOI-54 primary
    model. The dashed vertical lines correspond to the resonances
    shown in Fig.~\ref{f:resp-func}. The upper axis scale shows the
    tidal forcing angular frequency $\sigmk$
    (equation~\ref{e:sigmk}).}
  \label{f:gamma}
\end{figure}

Based on the reasoning in the preceding section, we make the ansatz
that the MD approach can be fixed by replacing the profile functions~(\ref{e:profile}) with
\begin{equation} \label{e:profile-rev}
  \profile[n,\ell,m,k]= \frac{\sigmk^{2}}{(\hsignl^{2} - \sigmk^{2}) - 2 \ii \gamlmk \, \sigmk},
\end{equation}
where the damping coefficient $\gamlmk$ does not depend on $n$. To
evaluate this coefficient we leverage energy conservation to
write\footnote{An equivalent relation also applies to the
stretched-string problem.}
\begin{equation} \label{e:gamlmk}
  \gamlmk = \frac{\work[\ell,m,k]}{2 \engy[\ell,m,k]}.
\end{equation}
Here
\begin{multline}
  \engy[\ell,m,k] = \sigmk^{2} \int_{0}^{R} \left[
    \left|\txir[\ell,m,k]\right|^{2} \right. \\
    \mbox{} +
    \left. \ell(\ell+1) \left|\txih[\ell,m,k]\right|^{2}
    \right] \rho r^{2} \, \diff{r}
\end{multline}
is the orbit-averaged total energy of the response to the
partial tidal potential $\pottide[\ell,m,k]$, and
\begin{multline}
  \work[\ell,m,k] = 2 \frac{G \Mpri^{2}}{a} q^{2}
  \left( \frac{\Rpri}{a} \right)^{\ell+3} \,
  \left( \frac{\rs}{\Rpri} \right)^{\ell+1} \\
  \times \imag(\Fbar[\ell,m,k]) \,
  \sigmk \,
  \Gbar[\ell,m,k]{5}
\end{multline}
is the orbit-averaged rate of work done on the star by this potential, with
\begin{equation}
  \Gbar[\ell,m,k]{5} \equiv \frac{2\ell+1}{4\pi} \left( \frac{R}{a} \right)^{-\ell+2} \left| \cbar[\ell,m,k] \right|^{2}
\end{equation}
(this definition means that $\Gbar[\ell,m,k]{4} = m \Gbar[\ell,m,k]{5}$).
The factor of 2 in the denominator of equation~(\ref{e:gamlmk})
accounts for the fact that energy scales as the square of the response
amplitude.

Because these expressions presume that solutions to the tidal
equations are already known, their usefulness in stand-alone MD
implementations may be limited. However, in the present case we can
bootstrap their evaluation using solutions from DS
approach. Fig.~\ref{f:gamma} plots the damping coefficient as a
function of rotation angular frequency, evaluated from DS solutions
for the same $\{\ell,m,k\}=\{2,-2,50\}$ partial tidal potential
considered previously. The coefficient varies significantly with
rotation frequency, with distinct peaks that correspond approximately
to the resonances seen in Fig.~\ref{f:resp-func}.

\begin{figure}
  \includegraphics{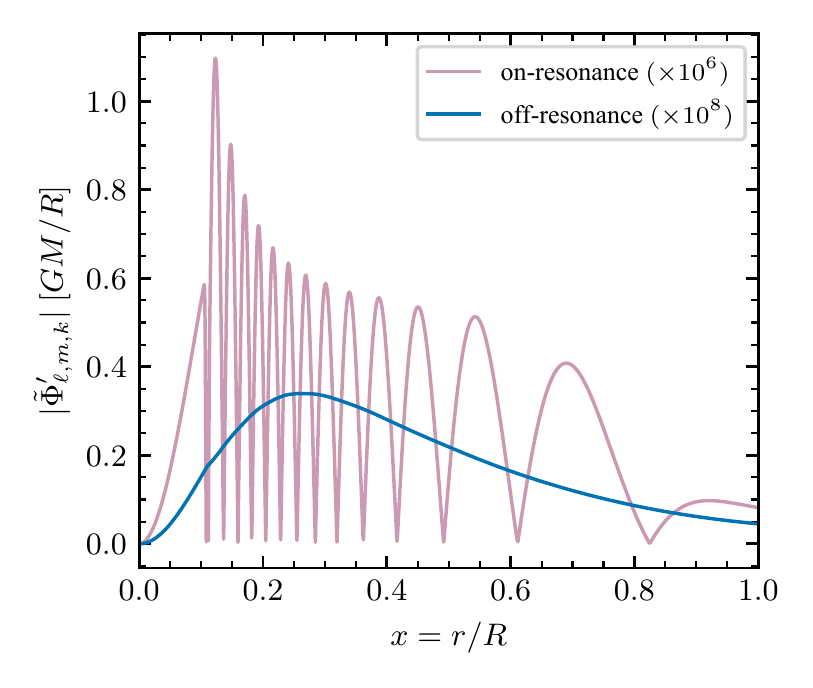}
  \caption{Eulerian self-gravitation potential perturbation
    $\tpotpri[\ell,m,k]'$ plotted against normalized radial coordinate $x$ for forcing on and off resonance with the $\gmode{15}$ mode.} 
  \label{f:wavefuncs}
\end{figure}

This behavior stems from the sensitivity of radiative dissipation to
the wavelength of perturbations. Figure~\ref{f:wavefuncs} plots the
self-gravitational potential perturbation as a function of radial
coordinate, obtained from the DS solutions at a pair of rotation
frequencies: one tuned to resonance with the $\gmode{15}$ mode, and
the other equidistant between the $\gmode{15}$ and $\gmode{16}$
resonances. The on-resonance case shows the highly oscillatory,
short-wavelength perturbations that are characteristic of dynamical
tides, and is subject to strong radiative damping: $\gamlmk/(2\pi) =
\SI{1.5e-6}{\per\day}$. In contrast, the off-resonance case shows
smooth, long-wavelength perturbations characteristic of equilibrium
tides, and the damping is reduced to $\gamlmk/(2\pi) =
\SI{3.4e-7}{\per\day}$.

\begin{figure}
  \includegraphics{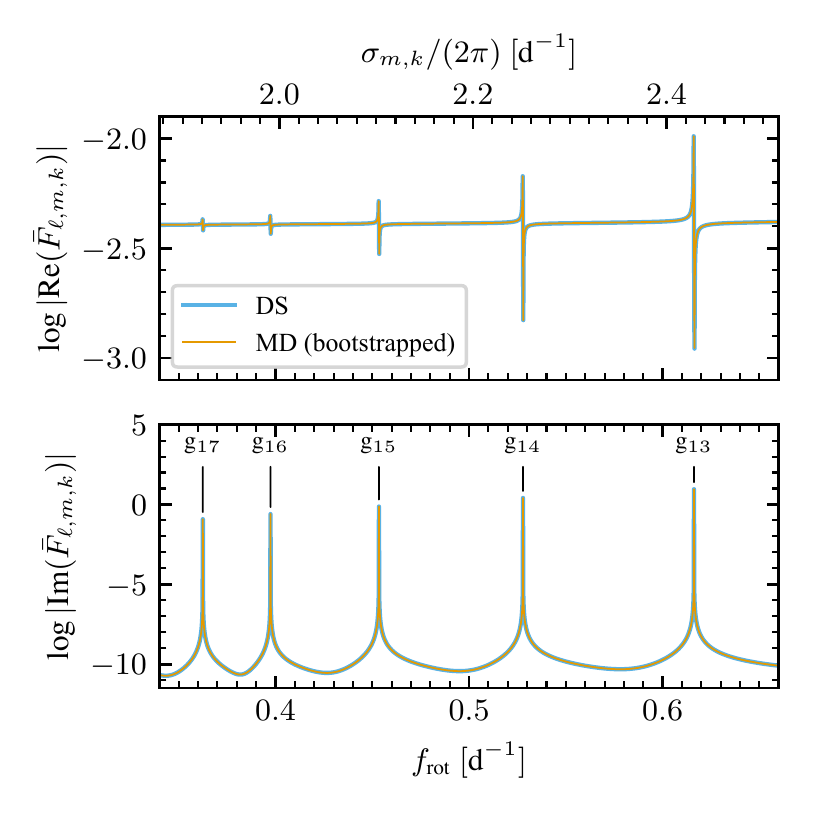}
  \caption{As in Fig.~\ref{f:resp-func}, except the MD approach uses
    the revised profile function~(\ref{e:profile-rev}) and damping
    coefficients bootstrapped using DS solutions (see
    Fig.~\ref{f:gamma}).}    
  \label{f:resp-func-bootstrap}
\end{figure}

\begin{figure}
  \includegraphics{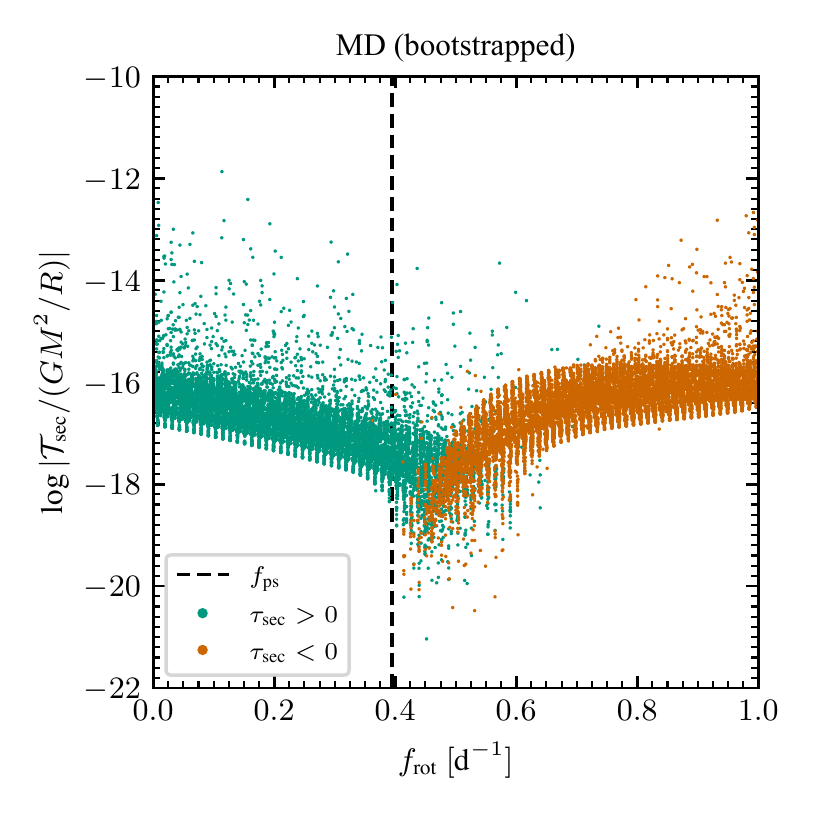}
  \caption{As in the right panel of Fig.~\ref{f:resp-func},
    except using the revised profile function~(\ref{e:profile-rev})
    and damping coefficients bootstrapped using DS solutions (see
    Fig.~\ref{f:gamma}).}
  \label{f:torque-bootstrap}
\end{figure}

With $\gamlmk$ determined, Fig.~\ref{f:resp-func-bootstrap} reprises
Fig.~\ref{f:resp-func} using the revised profile
functions~(\ref{e:profile-rev}) in the MD approach. The agreement
between DS and MD is greatly improved. As a further check,
Fig.~\ref{f:torque-bootstrap} reprises the right (MD) panel of
Fig.~\ref{f:torque}, but again using the revised profile
functions. There is now qualitative agreement between the two
approaches, with MD showing the same blurred transition from positive to
negative torque over a range centered on $\frot \approx
\SI{0.5}{\per\day}$, as does DS. However, noticeable differences still
remain; for instance, the MD torque values exhibit a lower envelope
that's around half an order of magnitude smaller than the DS ones,
and they also show significantly less scatter.

These lingering differences may stem from terms in
equation~(\ref{e:coeff}) other than $\profile[n,\ell,m,k]$; for
instance, as remarked by S23, numerical estimates for the overlap
integrals $Q_{n,\ell}$ can become inaccurate toward high radial
orders. However, as an alternative explanation we draw attention to a
more-fundamental limitation of the MD approach. In the stretched
string problem we assume the damping coefficient $\gamma$ is spatially
constant, and so the damping terms occur in the temporal
equation~(\ref{e:harm-osc}). This assumption cannot be made in the
stellar tides problem, because the radiative diffusivity varies by
many orders of magnitude throughout a star. Therefore, when solutions
of the form~(\ref{e:sol-xi},\ref{e:sol-f}) are adopted, the damping
terms instead appear in the spatial equations (see, e.g., Appendix E
of S23). The adiabatic eigenfunctions are not solutions to these
equations (a fact often overlooked in the literature), and the
MD approach loses its formal coherence.


\section{The Blurring of Pseudosynchronization} \label{s:pseudo}

Regardless of their remaining differences, Fig.~\ref{f:torque} (left)
and Fig.~\ref{f:torque-bootstrap} both agree that
pseudosynchronization is blurred in the KOI-54 primary model. While we
cannot offer a clear narrative for the characteristics of the blurring
(e.g., why in this case it manifests over the range $1.1 \la
\frot/\fps \la 1.4$), it is straightforward to explain why
equation~(\ref{e:pseudo}) does not apply. \citet{Hut:1981} derived
this equation within the weak friction approximation formalized by
\citet{Alexander:1973}, which presumes that the phase lag $\delta
\phi$ between the tidal forcing and the response is proportional to
$\sigmk$. Within the MD approach with the revised profile
function~(\ref{e:profile-rev}) the phase lag is given by
\begin{equation}
  \delta\phi = \arg(\profile[n,\ell,m,k]) = \tan^{-1} \left( \frac{2 \gamlmk \sigmk}{\hsignl^{2} - \sigmk^{2}} \right).
\end{equation}
Assuming $\gamlmk$ and $\sigmk$ are both much smaller than $\hsignl$
(in accordance with the presumption of a weakly damped equilibrium
tide), this approximates to
\begin{equation}
  \delta\phi \approx \frac{2 \gamlmk \, \sigmk}{\hsignl^{2}}
\end{equation}
When $\gamlmk$ itself depends on $\sigmk$ (see
Fig.~\ref{f:gamma}, in particular the upper axis scale), it's clear
that strict proportionality between $\delta\phi$ and $\sigmk$ cannot
hold; therefore, the weak friction approximation does not apply, and the derivation of equation~(\ref{e:pseudo}) breaks down.

If the original expression~(\ref{e:profile}) for the profile functions is used instead of the revised one~(\ref{e:profile-rev}), the phase lag becomes
\begin{equation}
  \delta\phi \approx \frac{2 \hgamnl \, \sigmk}{\hsignl^{2}},
\end{equation}
which \emph{is} proportional to $\sigmk$. This explains why \citet{Burkart:2012} were
able to reproduce equation~(\ref{e:pseudo}) from their analysis (see
their Appendix C2), and why their Fig.~4 and our Fig.~\ref{f:torque}
(right) show the equilibrium torque passing through zero at $\frot=\fps$.


\section{Summary \& Discussion} \label{s:discuss}

In the preceding sections we have demonstrated that a major source of
disagreement between the DS and MD approaches is an incorrect damping
coefficient in the MD profile functions~(\ref{e:profile}). With the
revised profile functions~(\ref{e:profile-rev}) the two approaches are
brought into closer although still incomplete agreement. The practical
issue of calculating damping coefficients $\gamlmk$ consistent with
energy conservation, without needing to bootstrap from DS solutions,
remains to be solved. One possible approach is to use a
quasi-adiabatic treatment that estimates damping coefficients from an
integral over the adiabatic tidal response --- for instance,
equation~(14) of \citet{Burkart:2012} modified to use the tidal
response wavefunctions $\txir[\ell,m,k]$, $\txih[\ell,m,k]$ rather
than individual modal eigenfunctions.

For stars like KOI-54 that have a convective core and a radiative
envelope, dynamical tides arising through modal resonances are
expected to be the dominant driver of orbital evolution
\citep{Zahn:1975}. This might imply that the highlighted issues with
MD are of little consequence: as Fig.~\ref{f:resp-func} shows, the MD
and DS tidal response functions are in good agreement at the resonance
peaks. However, the behavior of the tidal response in the
\emph{approach} to resonance peaks can be just as important as the
peaks themselves in determining the evolutionary trajectory of a
system. Consider for instance the phenomenon of resonance locking
\citep{Witte:1999}, where orbital and stellar evolution conspire to
maintain the frequency detuning $\delta \sigma = \hsignl - \sigmk$ at
a constant, small value over an extended period of time. The
existence, location and stability of a resonance lock depends on the
shape of the tidal response function in the wings of a peak
\citep[see, e.g., Fig.~1 of][]{Burkart:2014} --- exactly where the
deviations between MD and DS arise.

Although our analysis has focused mostly on the presentation of MD by
\citet{Kumar:1995} and \citet{Burkart:2012}, it extends to the broader
variety of decomposition approaches --- for instance, the phase-space
mode expansion favored by \citet{Schenk:2001} and \citet{Fuller:2017}.
Likewise, although we focus on stars with convective cores and
radiative envelopes, our findings are relevant for systems with other
internal architectures and other damping mechanisms. In any type of
star where the damping coefficient explicitly or implicitly depends on
the forcing frequency $\sigmk$, switching to the revised profile
functions~(\ref{e:profile-rev}) will alter the outcomes of MD-based
calculations, and potentially uncover a similar blurring of
pseudosynchronization. With this in mind, a careful re-evaluation of
results in the existing stellar tides literature seems prudent.

While pseudosynchronization does not operate in KOI-54 in the manner
originally envisaged by \citet{Hut:1981}, Figs.~\ref{f:torque} (left)
and~(\ref{f:torque-bootstrap}) nevertheless suggest that rotation
frequencies of systems like KOI-54 will tend to accumulate in the
range $1.1$--$1.4\,\fps$. This contrasts with the observational study of
24 heartbeat stars by \citet{Zimmerman:2017}, who find a majority of
systems rotate at $\sim 0.66\,\fps$. Resolving this tension between
theory and observation will be a productive direction for future
investigations.


\section*{Acknowledgments}

This work has been supported by NSF grants ACI-1663696, AST-1716436
and PHY-1748958, and NASA grant 80NSSC20K0515.

\facilities{We have made extensive use of NASA's
  Astrophysics Data System Bibliographic Services.}

\software{Astropy \citep{astropy:2013,astropy:2018,astropy:2022},
  \gyre\ \citep{Townsend:2013,Townsend:2018,Goldstein:2020,Sun:2023},
  Matplotlib \citep{Hunter:2007}, \mesa\ 
  \citep{Paxton:2011,Paxton:2013,Paxton:2015,Paxton:2018,Paxton:2019,Jermyn:2022}}


\bibliography{modal-decomp}

\end{document}